\documentclass[11pt,a4paper]{article}
\pdfoutput=1
\usepackage{jcappub}
\usepackage{bm}
\usepackage{enumitem}
\usepackage{braket}
\usepackage{mathrsfs}
\usepackage{slashed}
\usepackage{ulem}
\usepackage{braket}

\def\tot{{\rm tot}}

\def\Mpl{M_{\rm Pl}}

\def\eff{{\rm eff}}

\def\0{{(0)}}
\def\sig0{\dot{\sigma}_0}

\def\ph0{\dot{\phi}_0}

\def\att{\rm att}
\usepackage{color}
	\makeindex
		\title{
		When does the Schwinger Preheating Occur?}
		\author[a]{So Okano}
		\author[b,c]{and Tomohiro Fujita}
		
		\affiliation[a]{Department of Physics, Tokyo Institute of Technology,
			2-12-1 Ookayama, Meguro-ku, Tokyo 152-8551, Japan
		}
		\affiliation[b]{Waseda Institute for Advanced Study, Waseda University,
		1-6-1 Nishi-Waseda, Shinjuku, Tokyo 169-8050, Japan}
		\affiliation[c]{Research Center for the Early Universe (RESCEU), Graduate School of Science, The University of Tokyo, Hongo 7-3-1 Bunkyo-ku, Tokyo 113-0033, Japan}
		\emailAdd{okano.s.ab@m.titech.ac.jp}
		\emailAdd{tomofuji@aoni.waseda.jp}
		\abstract{When the inflaton couples to photons and amplifies electric fields, charged particles produced via the Schwinger effect can dominate the universe after inflation, which is dubbed as the Schwinger preheating. Using the hydrodynamic approach for the Boltzmann equation, we numerically study two cases, the Starobinsky inflation model with the kinetic coupling and the anisotropic inflation model.
		The Schwinger preheating is not observed in the latter model but occurs for a sufficiently large inflaton-photon coupling in the first model. We analytically address its condition and derive a general attractor solution of the electric fields. The occurrence of the Schwinger preheating in the first model is determined by whether the electric fields enter the attractor solution during inflation or not.
		}
	\arxivnumber{2105.xxxxx}
	
\begin{document}
		\maketitle
		\section{Introduction}
		\label{sec:intro}
The Schwinger effect, which is the non-perturbative pair production in the strong electromagnetic fields, 
has been theoretically predicted~\cite{Sauter:1931,Heisenberg:1936,Schwinger:1951nm}. The reaction rate in the case of homogeneous and constant electric fields, $E$, is proportional to $\exp\left(-\pi \frac{m^2}{eE}\right)$, where $m$ and $e$ represent the mass and the coupling constant of the produced particles, respectively. 
The critical strength of electric fields is given by $E_{\rm cr}=\pi m^2/e \sim 10^{18}\,{\rm V/m}$ for the electrons and the positrons, and
it has not yet been reached by experiments. 
However, the European Extreme Light Infrastructure (ELI) project and the X-ray free electron laser (XFEL) project aim to 
achieve $E\sim 10^{-1}E_{\rm cr}$ ~\cite{Dunne:2008kc,Ringwald:2001ib}, and therefore the Schwinger effect would be experimentally verified in the near future. 

Several physical situations which may cause the Schwinger effect are discussed, such as relativistic heavy ion collisions~\cite{Casher:1978wy}, neutron stars~\cite{Ruffini:2003cr,Yatabe:2017zql}, charged black holes~\cite{Gibbons:1975kk,Chen:2012zn}, and inflation with the amplification of the electromagnetic fields~\cite{Turner:1987bw,Ratra:1991bn,Garretson:1992vt,Field:1998hi,Bamba:2003av,Anber:2006xt,Caprini:2014mja,Kobayashi:2014sga,Fujita:2015iga,Fujita:2016qab,Vilchinskii:2017qul,Caprini:2017vnn,Fujita:2019pmi,Okano:2020uyr}. In particular, such inflation models are 
intensively studied in the context of magnetogenesis and baryogenesis. Since the production of the charged particles affects the evolution of the electromagnetic fields, the Schwinger effect should be taken into account in the analysis. However, 
it is a complicated task and various approaches have been discussed
~\cite{Garriga:1994bm,Martin:2007bw,Frob:2014zka,Cai:2014qba,Kobayashi:2014zza,Stahl:2015gaa,Bavarsad:2016cxh,Hayashinaka:2016qqn,Domcke:2018gfr,Kitamoto:2018htg,Domcke:2019qmm,Sobol:2020frh,Kitamoto:2020tjm,Domcke:2021fee}. The difficulties of the analysis arise from the time dependence of the electric fields and the curved spacetime. In fact, even in the flat spacetime, the Schwinger effect in general dynamical electric fields is difficult to analyze because we cannot analytically obtain the mode functions of the charged particles, and thus we need some prescriptions. One of the well studied prescriptions for the uniform and dynamical electromagnetic fields is the quantum kinetic approach~\cite{Rau:1994ee,Schmidt:1998vi,Kluger:1998bm,Bloch:1999eu,Dumlu:2009rr,Kim:2011jw,Huet:2014mta,Gorbar:2019fpj}, where the time evolution of the momentum distribution function of the charged particles is described by the quantum Boltzmann equation. 

Ref.~\cite{Gorbar:2019fpj} approximately applied the quantum kinetic approach to the kinetic coupling inflation model without magnetic fields. In this model, the inflaton is coupled to the electromagnetic fields through the gauge kinetic function, $f(\phi)^2F^{\mu\nu}F_{\mu\nu}/4$~\cite{Ratra:1991bn}. Ref.~\cite{Gorbar:2019fpj} found that the charged particles produced by the Schwinger effect temporarily dominate the universe right after inflation, which they call ``Schwinger preheating''. 
This is a new fascinating phenomenon and is worth a closer investigation. 
In addition, the oscillation of the electric fields after inflation was found.
With a large amount of charged particles, the electric fields are naively expected to exponentially decay due to the electric conductivity.
In their analysis, however, the electric fields do not quickly decay but persistently oscillate with the charged particles. 
These findings indicate the importance of the Schwinger effect in the inflation models with the electromagnetic fields.  
Nonetheless, they studied only one inflation model and did not give the condition for the Schwinger preheating. 
The calculation in Ref.~\cite{Gorbar:2019fpj} also assumed the isotropic background spacetime, which is inconsistent with
the existence of the background electric field.
When the Schwinger preheating occurs, the energy density of the electric fields become comparable to the total energy density at the end of the inflation, 
and the spacetime anisotropy may not be negligible.
Thus, the generalities of the Schwinger preheating as well as the electric field oscillation are still unclear.

In this paper, we numerically scrutinize two inflaton models. One is the same model as Ref.~\cite{Gorbar:2019fpj} and the other is the anisotropic inflation model~\cite{Watanabe:2009ct}.
With analytic arguments, we derive the attractor solution of the electric field and find the general condition for the Schwinger preheating. 
Furthermore, we take into account the anisotropy of the spacetime and estimate its impact on the dynamics.  
Our analysis implies that the oscillation of the electric fields can occur independently from the model of inflation, but the Schwinger preheating does not always occur.

This paper is organized as follows. In section~\ref{sec:kinetic}, we introduce the formulation of the kinetic approach to the inflationary universe and present some numerical results. In section~\ref{sec:preheating}, we derive the attractor solution for the electric fields and give the condition for the Schwinger preheating. 
The influence of the spacetime anisotropy is discussed in section~\ref{sec:anisotropy}. Finally, section~\ref{sec:Summary and Discussion} is devoted to the summary and discussion.

\section{Kinetic approach to the Schwinger effect in the early universe}
\label{sec:kinetic}
		In this section, we study the dynamics of the charged particles produced via the Schwinger effect by using Boltzmann equation for one particle distribution function in the inflationary background and the subsequent reheating era. We first review the inflation dynamics in an anisotropic background induced by electric fields. Then we introduce the Boltzmann equation in the anisotropic background and give the simple representations of the source term and the collision term by following the procedure of Ref.~\cite{Gorbar:2019fpj}. We introduce the moment of the distribution function and reduce the Boltzmann equation into the hydrodynamic equations. Finally we show the numerical results of two models in the case of the isotropic background. In Sec.~\ref{sec:anisotropy}, we will discuss the influence of background anisotropy on the numerical results which we neglect in this section.
		
		\subsection{Inflation in the anisotropic background}
		\label{sec:anisotropic inflation}
		
		Here we briefly review the inflation dynamics in the anistropic background and derive the background equations for the spacetime, inflaton, and electric fields. Our Lagrangian for the gravity, inflaton, electromagnetic fields, and charged particles are given by
			\begin{align}
			\mathcal{L}=-\frac{\Mpl^2}{2}\mathcal{R}+\frac{1}{2}\left(\partial_{\mu}\phi\right)^2-V(\phi)-\frac{1}{4}f(\phi)^2F_{\mu\nu}F^{\mu\nu}+\mathcal{L}_{\psi}.
			\end{align}
			Here $\mathcal{R},~\phi,~F_{\mu\nu}=\partial_{\mu}A_{\nu}-\partial_{\nu}A_{\mu}$ are the Ricci scalar, the inflaton, the field strength of the $U(1)$  gauge field $A_{\mu}$, respectively. $V(\phi)$ is the potential of the inflaton, and $ f(\phi)$ is the kinetic coupling function of the inflaton to the $U(1)$ gauge fields, which will be specified later. $\Mpl$ is the reduced Planck mass. $\mathcal{L}_{\psi}$ is the lagrangian of charged particles interacting with electromagnetic fields. To be concrete, we consider the Dirac fermions $\mathcal{L}_{\psi}=\bar{\psi}(i\slashed{D}-m) \psi$, while the extension to  charged scalars is straightforward. Without loss of generality, we can choose the $x$-axis as the direction of the electric fields. 
			For simplicity, we assume that the direction of the electric fields does not change in time. In the homogeneous universe, we consider $\phi=\phi(t)$ and $A_{\mu}=\left(0,A(t),0,0\right)$, where we used the gauge fixing condition $A_0=0$ due to the gauge invariance. 
			Note that the magnetic fields are neglected in this treatment. The consistent background spacetime metric for this field configuration is known as the Bianchi-I spacetime,
			\begin{align}
			\label{eq:Bianchi-I}
			ds^2=dt^2-e^{2\alpha(t)}\left(e^{-4\sigma(t)}dx^2+e^{2\sigma(t)}(dy^2+dz^2)\right).
			\end{align}
			Here  $\alpha(t)$ and $\sigma(t)$ represent the e-foldings of the isotropic scale factor and deviation from the isotropy respectively.  
			We solve the following evolution euqations of the spacetime $\alpha(t), \sigma(t)$, 
			the inflaton $\phi(t)$ 
			and the electric energy density $\rho_E(t)$:
			\begin{align}
			\dot{\alpha}^2-\dot{\sigma}^2 &=\frac{1}{3\Mpl^2}\left(\rho_{E}+\rho_{\psi}+V(\phi)+\frac{1}{2}\dot{\phi}^2\right),
			\\
			\ddot{\sigma}+3\dot{\alpha}\dot{\sigma} &=\frac{2\rho_E}{3\Mpl^2},
			\\
			\ddot{\phi}+3\dot{\alpha}\dot{\phi}+V'(\phi)&=2\frac{f'}{f}\rho_E,
			\label{eq:EoM for istoropic phi}
			\\
			\dot{\rho}_E+4\left(\dot{\alpha}+\dot{\sigma}\right)\rho_E &=-2\frac{f'}{f}\dot{\phi}\rho_E-e^{\alpha} \tilde{j}_{\rm tot}\mathcal{E}
			+\frac{\dot{\alpha}^3}{4\pi^2}\left[\dot{\alpha}^2+\left(\frac{\dot{f}}{f}\right)^2\right],
			\label{eq:EoM for rhoe an}
			\end{align}
            where dot denotes the cosmic time derivative and prime denotes the inflaton derivative, e.g. $f'\equiv \partial_\phi f(\phi)$.
			The physical electric fields are defined as $E(t)=-e^{-\alpha+2\sigma}\dot{A}$ and its energy density is $\rho_E=\frac{1}{2}f^2E^2$.
			$\rho_{\psi}$ represents the energy density of the charged particles. 
			$\tilde{\bm{j}}_{\rm tot}$ is the total electric current of the charged particles, which flows along the electric fields $\bm{E}$. 
			We used $e\tilde{\bm{j}}_{\rm tot}\cdot \bm{E} = \tilde{j}_{\rm tot} \mathcal{E}$ by introducing $\mathcal{E}\equiv eE(t)$, where $e$ is the coupling constant of the fermion with the electric fields.	The last term in the right hand side of Eq.~\eqref{eq:EoM for rhoe an} is phenomenologically added. 
			With the varying kinetic function $f$, the fluctuation of the gauge field is produced in the inflationary background
			and its contribution to the background electric energy is roughly estimated as~\cite{Sobol:2018djj}
			\begin{align}
			(\dot{\rho}_E)_H=\left.\frac{d\rho_E}{dk}\right|_{k=k_H}\cdot\frac{dk_{H}}{dt}\simeq \frac{\dot{\alpha}^3}{4\pi^2}\left[\dot{\alpha}^2+\left(\frac{\dot{f}}{f}\right)^2\right].
			\end{align}
			This contribution gives a finite initial amplitude for the electric field which is subsequently amplified by the inflaton. Since the electric field is eventually stabilized at its attractor value, the initial amplitude is not very important and we adopt the same expression for $(\dot{\rho}_E)_H$ as Ref.~\cite{Sobol:2018djj}.
		The drawback of Eq.~\eqref{eq:EoM for rhoe an} is that we cannot track the direction of the electric field.
		Therefore, when we numerically solve the set of the equations, we use not Eq.~\eqref{eq:EoM for rhoe an} but the EoM for electric fields $\mathcal{E}=ef^{-1}\sqrt{2\rho_E}$,%
		\footnote{In our numerical calculation, we multiply the last term by the Heaviside step function $\theta (\mathcal{E}-10^{-100}\mu^2)$ to avoid the singularity at $\mathcal{E}=0$.}
		\begin{align}
		\dot{\mathcal{E}}+2(\dot{\alpha}+\dot{\sigma})\mathcal{E}+2\frac{\dot{f}}{{f}}\mathcal{E}=-\frac{e^2}{f^2} j_{\rm tot}+\frac{e^2}{f^2}\frac{\dot{\alpha}^3}{4\pi^2\mathcal{E}}\left[\dot{\alpha}^2+\left(\frac{\dot{f}}{f}\right)^2\right],
		\end{align}
		where we introduce $j_{\rm tot}\equiv e^{\alpha}\tilde{j}_{\rm tot}$, which is the electric current density for the comoving observer.
	
	    A well-known example of inflation accompanied by significant electric fields is the anisotropic inflation model~\cite{Watanabe:2009ct}.
		In this model, the inflaton potential and the kinetic coupling function are given by
		\begin{align}
		\label{eq:model of anisotropic inflation}
		V_W(\phi)=\frac{1}{2}\mu^2\phi^2, \qquad f_W(\phi)=\exp\left(\frac{c\phi^2}{2\Mpl^2}\right),
		\end{align}
		where $\mu$ is the mass of inflaton, and $c$ is a dimensionless coupling constant. 
		When we ignore the charged particles, this system has an attractor solution of the electric field~\cite{Watanabe:2009ct},
		\begin{align}
		\rho_E^{\att}=\frac{3}{2}\frac{c-1}{c^2}\epsilon_V\Mpl^2\dot{\alpha}^2,
		\label{eq:rhoE att Watanabe}
		\end{align}
		with $\epsilon_V=\frac{\Mpl^2}{2}\left(\frac{V'}{V}\right)^2$. 
		Note that we should set $c>1$ to have a non-vanishing value of the electric field in the attractor solution.
		
		\subsection{Boltzmann equation in the anisotropic background}
		\label{sec:Hydrodynamic approach to the Boltzmann equation}
		We employ the Boltzmann equation in the curved spacetime with the electromagnetic fields to describe the dynamics of the charged particles,
		\begin{align}
		\label{eq:Boltzmann1}
		\left[P^{\mu}\partial_{\mu}-\Gamma_{\nu\lambda}^{\mu}P^{\nu}P^{\lambda}\frac{\partial}{\partial P^{\mu}}-e P^{\mu}F_{\mu j}\frac{\partial}{\partial P_j}\right]\mathcal{F}(x,P)=u_{\mu}P^{\mu}\left(\mathcal{S}[\mathcal{F}]+C[\mathcal{F}]\right),
		\end{align}
		where $u, P, \mathcal{S}, \mathcal{C}$ and $\mathcal{F}(x,P)$ represent the four-velocity of the local flow, the four-momentum of the particle, the Schwinger production term, the collision term, and the distribution function, respectively. $\Gamma^\mu_{\nu\lambda}$ denotes the Levi-Civita connetiction.
		When we consider the anisotropic background, Eq.~\eqref{eq:Bianchi-I}, and the electric fields in the $x$-direction, the Boltzmann equation can be rewritten as
		\begin{align}
		\label{eq:Boltzmann an}
		\frac{\partial \mathcal{F}_p}{\partial t}+\left(\mathcal{E}(t) +3\dot{\sigma}p_{\parallel}\right)\frac{\partial \mathcal{F}_p}{\partial p_{\parallel}}-(\dot{\alpha}+\dot{\sigma})\bm{p}\cdot\frac{\partial \mathcal{F}_p}{\partial \bm{p}}=\mathcal{S}[\mathcal{F}_p]+\mathcal{C}[F_{p}],
		\end{align}
		 where we introduced the physical momentum $\bm{p}$ with $P^{\mu}=(\epsilon_{\bm{p}},\ e^{-\alpha+2\sigma}\bm{p}_{\parallel},\  e^{-\alpha-\sigma}\bm{p}_{\perp})$ and $\bm{p}^2=\bm{p}_{\parallel}^2+\bm{p}_{\perp}^2.$
		 One can reproduce the equation in the isotropic background by taking the limit $\dot{\sigma}\rightarrow 0$.
		 
		 To solve the Boltzmann equation, we need to specify the collision term and the Schwinger source term. In this part, we basically follow the procedure used in the previous work~\cite{Gorbar:2019fpj}. We assume that
		 the collision term can be approximated by a restoring force towards the thermal equilibrium with a relaxation time $\tau$,
		\begin{align}
		\mathcal{C}[\mathcal{F}]=-\frac{\mathcal{F}_{\bm{p}}-\mathcal{F}_{\bm{p}}^{\rm eq}}{\tau}
		\end{align}
		where $\mathcal{F}_{p}^{\rm eq}$ represents the Fermi-Dirac distribution function without chemical potential $\mathcal{F}_{p}^{\rm eq}=\left(\exp\left[(u^{\nu}P_{\nu}/T)\right]+1\right)^{-1}$ with $T=[60 \rho_\psi/(7\pi^2)]^{1/4}$.
        It is known that the relaxation time of the electron in the flat spacetime is $\tau=\frac{c_0(4\pi)^2}{e^4T\ln e^{-1}}$, where $c_0$ is an order 1 parameter~\cite{Thoma:2008my}. We adjust this expression to our setup. $e$ should be replaced by the effective coupling $e_{\eff}=e/f$. The temperature $T$ in $\tau$ is approximated by the average kinetic energy of the electron, $\rho_{\psi}/n$. Finally we chose $c_0=1$. Then we obtain the relaxation time as
        \begin{align}
        \tau=\frac{(4\pi)^2}{\left(\frac{\rho_{\psi}(t)}{n(t)}\right)e_{\eff}^4\ln e_{\eff}^{-1}}.
        \end{align}

		The Schwinger source term in the kinetic approach in the inflationary universe was estimated as~\cite{Gorbar:2019fpj},
		\begin{align}
		\label{eq:Schwinger term}
		\mathcal{S}[f]=\left(1-2\mathcal{F}(t,\bm{p})\right)\sqrt{|\mathcal{E}(t)|}\exp\left(-\pi\frac{m^2+\bm{p}^2}{|\mathcal{E}(t)|}\right).
		\end{align}
		The Pauli blocking factor $(1-\mathcal{F})^2$ is expanded into $\left(1-2\mathcal{F}\right)$ in the leading order of $\mathcal{F}$. 
		One can show that Eq.~\eqref{eq:Schwinger term} reproduces the pair production rate for the constant electric fields,
 		\begin{align}
		\label{eq: reaction rate}
		\Gamma=\int \frac{d^3\bm{p}}{(2\pi)^3}\mathcal{S}[\mathcal{F}=0]=\frac{|\mathcal{E}|^2}{(2\pi)^3}\exp\left(-\frac{\pi m^2}{|\mathcal{E}|}\right),
		\end{align}
		where we 
		ignored the Pauli blocking factor. 
		We also find the polarization current density, $j_{\rm pol}$, of the produced charged particles which is not yet accelerated by the electric fields. In the strong field limit, $|\mathcal{E}|\gg H^2, m^2$, which is applicable to our case, $j_{\rm pol}$ is computed as
		\begin{align}
		\label{eq: polarization current}
		j_{\rm pol}=\frac{2g}{\mathcal{E}}\int \frac{d^3\bm{p}}{(2\pi)^3}\sqrt{m^2+\bm{p}^2}\mathcal{S}[\mathcal{F}=0]\simeq g\, {\rm sign}(\mathcal{E}) \frac{|\mathcal{E}|^{3/2}}{(2\pi)^3},
		\end{align}
		where $g=2s+1$ is the number of spin projection and $s=1/2$ in our case. Here we ignored $\mathcal{F}$ again.
		We distinguish $j_{\rm pol}$ from the conductive current, $j_{\rm cond}$, which is the current density of the charged particles accelerated by the electric fields after their production. When we consider the interaction between the electric fields and charged particles, we should include the both current densities, $j_{\tot} =j_{\rm pol}+j_{\rm cond}$.
        
        Our numerical calculations use Eqs.~\eqref{eq: reaction rate} and \eqref{eq: polarization current} which assume that the number of the charged particles per unit volume in the phase space is negligible, $\mathcal{F}\approx 0$.
        The approximation of the vanishing $\mathcal{F}$ may be justified, since the rapid expansion of the spacetime and the acceleration by the electric field change the momenta of the charged particles and empty the phase space that were filled up by the previous particle production.
        Note that Ref.~\cite{Gorbar:2019fpj} directly solved the Boltzmann equation without using this approximation $\mathcal{F}\approx 0$ and confirmed that the hydrodynamic approach reproduced the qualitatively equivalent results with this approximation. Thus we adopt the same approximation.
		 
		Let us introduce a $(k, r)$ moment to apply the hydrodynamic approach for Eq.~\eqref{eq:Boltzmann an} as,
		 \begin{align}
		 \mathcal{J}_{i_1,\cdots,i_k}^{(k, r)}(t)=2g\int \frac{d^3\bm{p}}{(2\pi)^3}\frac{p_{i_1}\cdots p_{i_{k}}}{\epsilon(\bm{p})^{r+2k-1}}\mathcal{F}(t,\bm{p}),
		 \end{align}
		 where the factor 2 appears from the contribution of the electron and positron.  
		 Multiplying the Boltzmann equation by $\frac{p_{i_1}\cdots p_{i_{k}}}{\epsilon(\bm{p})^{r+2k-1}}$ and integrating over the momentum space, we obtain the following chain of ordinary differential equations:
		\begin{align}
		\label{eq:hydro boltzmann an}
		&\left[\frac{d}{dt}+\frac{1}{\tau}-(r+k-4)(\dot{\alpha}+\dot{\sigma})-3\dot{\sigma}\right]\mathcal{J}_{i_1\cdots i_k}^{(k,r)}-3\dot{\sigma}\sum_{\ell=1}^{k}\mathcal{J}_{i_1\cdots \hat{i_{\ell}}\cdots i_k x}^{(k,r)}+3\dot{\sigma}(r+2k-1)\mathcal{J}_{i_1\cdots i_kxx}^{(k+2,r-2)}\notag\\
		&+\mathcal{E}\left[(r+2k-1)\mathcal{J}_{i_1\cdots i_k,x}^{(k+1,r)}-\sum_{\ell=1}^k\delta_{\ell,x}\mathcal{J}_{i_1\cdots \hat{i_{\ell}}\cdots i_k}^{(k-1,r+2)}\right]+(r+2k-1)(\dot{\alpha}+\dot{\sigma})m^2\mathcal{J}_{i_1\cdots i_k}^{(k,r+2)}\notag\\
		&=2g\int \frac{d^3\bm{p}}{(2\pi)^3}\frac{p_{i_1}p_{i_2}\dots p_{i_k}}{\epsilon(\bm{p})^{r+2k-1}}\mathcal{S}(\bm{p})+2g\int \frac{d^3\bm{p}}{(2\pi)^3}\frac{p_{i_1}p_{i_2}\dots p_{i_k}}{\epsilon(\bm{p})^{r+2k-1}}\mathcal{F}_{\rm eq}(\bm{p},T).
		\end{align}
		The detailed derivation is described in Appendix.~\ref{appendixA}.
		Here $\hat{i}_l$ means that the index $i_l$ is absent in the sequence. Since we consider that the charged particles are relativistic and $\mathcal{E}\gg m^2$, we ignore the contribution from the mass term in the above equation. We cannot solve the infinite chains of equations and hence we need closure conditions to cut off the chains. Now we extract the three pieces of the chains of equations for the $(0, 1), (0, 0), (1, 0)$ moments, which are important to describe the Schwinger pair production. We specify each moments as the energy density of the charged particles, $\mathcal{J}^{(0, 0)}\equiv \rho_{\psi}$, their number density, $\mathcal{J}^{(0, 1)}\equiv n$,  and their conductive current density along electric fields, $\mathcal{J}_{x}^{(1, 0)}=j_{\rm cond}$. The evolution equations for them are written as,
		\begin{align}
		\label{eq:eom for n}
		&\frac{dn}{dt}+\left(\frac{1}{\tau}+3\dot{\alpha}\right)n=2g\Gamma+\frac{n_{\rm eq}}{\tau},\\
		\label{eq:eom for rhoc}
		&\frac{d\rho_{\psi}}{dt}+\left(4\dot{\alpha}+\dot{\sigma}\right)\rho_{\psi}-3\dot{\sigma}\mathcal{J}_{xx}^{(2,-2)}=\mathcal{E}j_{\rm pol}+\mathcal{E}j_{\rm cond},\\
		\label{eq:eom for jcond}
		&\frac{dj_{\rm cond}}{dt}+\left(\frac{1}{\tau}+3\dot{\alpha}\right)j_{\rm cond}-3\dot{\sigma}j_{\rm cond}+3\dot{\sigma}\mathcal{J}_{x xx}^{(3,-2)}=\mathcal{E}\mathcal{J}_{\perp\perp}^{(2,0)},
		\end{align}
		where $n_{\rm eq}=2g\int \frac{d^3\bm{p}}{(2\pi)^3} \mathcal{F}_{\rm eq}$. The reaction rate $\Gamma$ and the polarization current $j_{\rm pol}$ are given by Eqs.~\eqref{eq: reaction rate} and \eqref{eq: polarization current}, respectively. 
		The variables whose expressions have not yet been specified are $\mathcal{J}_{xx}^{(2, -2)}$,  $\mathcal{J}_{xxx}^{(3, -2)}$ and $\mathcal{J}_{\perp\perp}^{(2, 0)}$.
		
		Here, for simplicity, we ignore the anistropy for a while. Taking the isotropic limit, $\dot{\sigma}\rightarrow 0$,
		one finds that $\mathcal{J}_{xx}^{(2, -2)}$ and $\mathcal{J}_{xxx}^{(3, -2)}$ become irrelevant to the evolution of the system.  
		$\mathcal{J}_{\perp\perp}^{(2, 0)}$ is the only variable remaining relevant but unspecified. 
		Note that we will restore the anisotropy and discuss its influence in Sec.~\ref{sec:anisotropy}.
		
		Let us identify $\mathcal{J}_{\perp\perp}^{(2, 0)}$ through a microscopic arguments. 
		Assuming the conductive current is approximated by the bulk flow of the charged particles in the $x$-direction, $j_{\rm cond}=n v_x$,
		we obtain the EoM for the charged particle by combining Eq.~\eqref{eq:eom for n} and Eq.~\eqref{eq:eom for jcond} with $\dot{\sigma}\rightarrow 0$ as,
		\begin{align}
		\label{eq: derudes1}
		\frac{d v_x}{dt}=-\frac{v_x}{n}\left(2g\Gamma+\frac{n_{\rm eq}}{\tau}\right)+\frac{\mathcal{E}}{n}\mathcal{J}_{\perp\perp}^{(2,0)}.
		\end{align}
		The first term in the R.H.S of the above equation describes the slowing down of the particle by creating new particles whose averaged velocity is zero and scattering with the other particles. The second term represents the Lorentz force from the electric fields. 
		The EoM for the relativistic charged particle in the isotropic background, $ds^2=dt^2-e^{2\alpha}\left(dx^2+dy^2+dz^2\right)$ under the electric field in the $x$-direction is written as,
		\begin{align}
		\label{eq: derudes2}
		\frac{dv_x}{dt}=-Cv_x+\mathcal{E}\frac{n}{\rho_{\psi}}(1-v_x^2),
		\end{align}
		where $C$ is the decay constant and we replaced the typical momentum of the particle $p$ by $\rho_\psi/n$ in the second term~\cite{Fujita:2019pmi}. Comparing Eq.~\eqref{eq: derudes1} with Eq.~\eqref{eq: derudes2}, we obtain the closure condition as,
		\begin{align}
		\mathcal{J}_{\perp\perp}^{(2,0)}&=\frac{n^2-j_{\rm cond}^2}{\rho_{\psi}}.
		\end{align}

 \subsection{Time evolution of the system}
 \label{sec:time evolution}

In this section, we will present the numerical results of the time evolution of the system. 
We solve the following EoMs for the inflaton, the spacetime, and electric fields in the isotropic background:
\begin{align}
\dot{\alpha}^2&=\frac{1}{3\Mpl^2}\left(\rho_{E}+\rho_{\psi}+V(\phi)+\frac{1}{2}\dot{\phi}^2\right),
\\
\ddot{\phi}&=-3\dot{\alpha}\dot{\phi}-V'(\phi)+2\frac{f'}{f}\rho_E,
\notag\\
\label{eq:EoM for rhoe}
\dot{\mathcal{E}}+2\dot{\alpha}\mathcal{E}+2\frac{\dot{f}}{{f}}\mathcal{E}&=-\frac{e^2}{f^2} j_{\rm tot}+\frac{e^2}{f^2}\frac{\dot{\alpha}^3}{4\pi^2\mathcal{E}}\left[\dot{\alpha}^2+\left(\frac{\dot{f}}{f}\right)^2\right].
\end{align}
We simultaneously solve the evolution of the hydrodynamic variables introduced in the previous section. The hydrodynamic equations are given by
\begin{align}
\frac{dn}{dt}+\left(\frac{1}{\tau}+3\dot{\alpha}\right)n&=2g\Gamma+\frac{n_{\rm eq}}{\tau},\\
\frac{d\rho_{\psi}}{dt}+4\dot{\alpha}\rho_{\psi}&=\mathcal{E}j_{\rm pol}+\mathcal{E}j_{\rm cond},\\
\frac{dj_{\rm cond}}{dt}+\left(\frac{1}{\tau}+3\dot{\alpha}\right)j_{\rm cond}&=\mathcal{E}\frac{n^2-j_{\rm cond}^2}{\rho_{\psi}}.
\end{align}
To fix its mass, we identify the fermions as electrons and positrons. Their mass is given by the Yukawa coupling and the Higgs field vev, $h(t)$, which obtains a large root mean square value due to the quantum fluctuations during inflation. The mass of the electron is estimated as
\begin{align}
m(t)=\frac{y_e}{\sqrt{2}}h(t)\simeq \frac{y_e}{\sqrt{2}}\lambda^{-1/4}\dot{\alpha}(t),
\end{align}
where $y_e=3\times 10^{-6}$ and $\lambda=0.26$. 
Note that when the particle production via the Schwinger effect is the most efficient,
this mass is negligibly smaller than the electric field, $m\ll \mathcal{E}$
and thus its precise value does not affect our final results.

\subsubsection{Starobinsky model with Ratra coupling}

\begin{figure}
			\centering	\includegraphics[width=.8\textwidth]{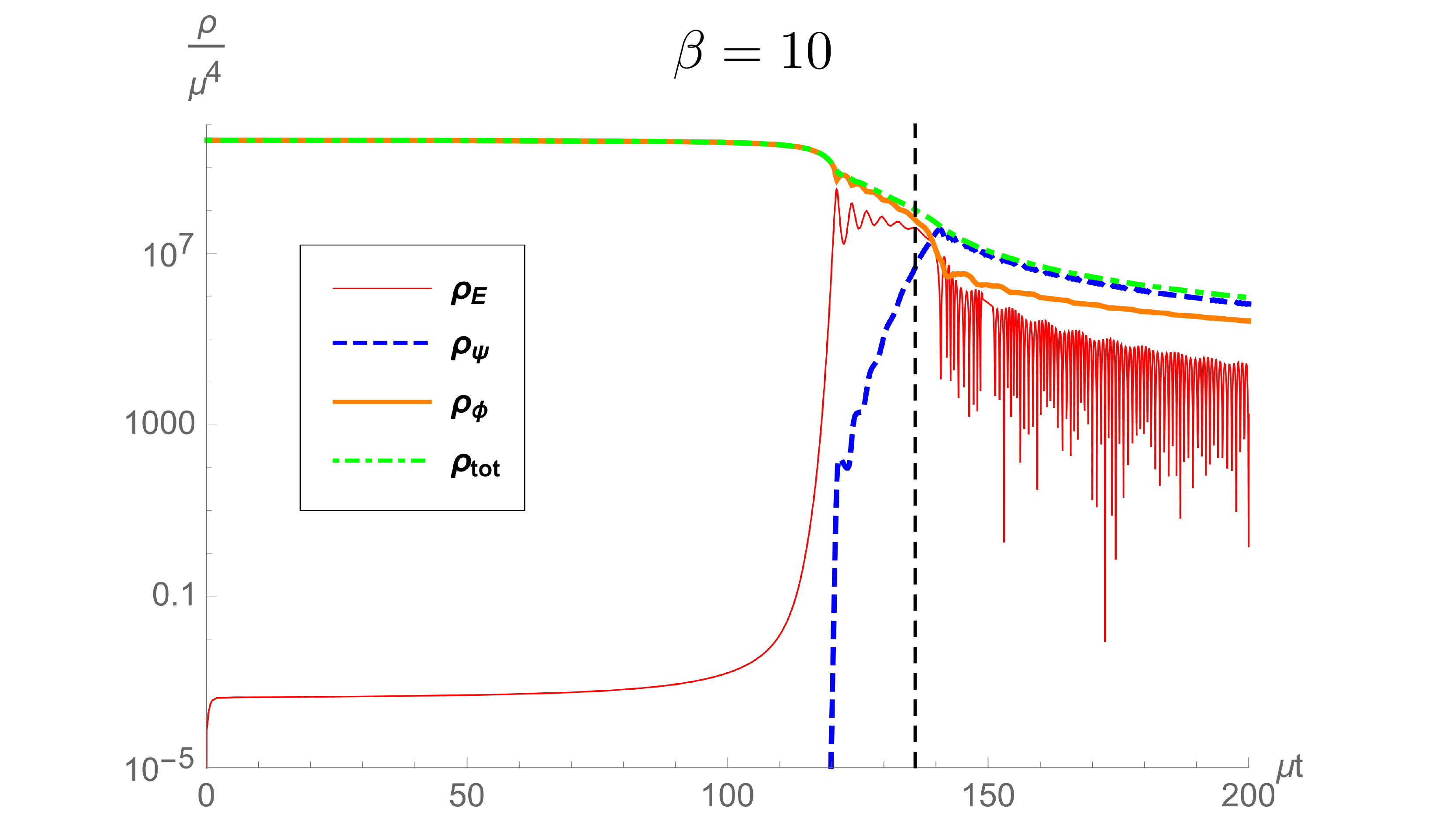}	
			\caption{ 
			The numerical result in case of the Starobinsky model $V_S$ and the Ratra coupling $f_R$ with the coupling constant $\beta=10$. We show the time evolution of the energy densities of the electric field $\rho_E$ (red solid), the charged fermions $\rho_\psi$ (blue dashed), the inflaton $\rho_\phi$ (orange solid) and their total $\rho_{\rm tot}$ (green dot-dashed). The calculation starts at $\mu t=0$ corresponding to $N\simeq 60$ and inflation ends at $\mu t=136
			$ (vertical black dashed line) where $\epsilon_H\equiv -\ddot{\alpha}/\dot{\alpha}^2=1$. 
			The produced fermions dominate the universe after inflation, $\rho_\psi> \rho_\phi,\rho_E$, which is called the Schwinger preheating.
			}
			\label{fig:sobolrhovst}
		\end{figure}

The first model that we study is the Starobinsky inflation model~\cite{Starobinsky:1980te} with the Ratra coupling function~\cite{Ratra:1991bn},
\begin{align}
\label{eq:starobinskyratra}
V_{\rm S}(\phi)=\frac{3\mu^2\Mpl^2}{4}\left[1-\exp\left(-\sqrt{\frac{2}{3}}\frac{\phi}{\Mpl}\right)\right]^2, \qquad
f_{R}(\phi)=\exp\left(\beta\frac{\phi}{\Mpl}\right),
\end{align}
where $\mu\simeq1.3\times 10^{-5}\Mpl$ is the mass of the inflaton around $\phi=0$.
To perform the numerical calculation, we set the initial condition at $N= 60$ based on
\begin{align}
\phi(0)=\sqrt{\frac{3}{2}}\Mpl\ln \left(\frac{4}{3}N\right),
\qquad
\phi'(0)=-\frac{\mu\Mpl}{2N}\sqrt{\frac{3}{2}},
\end{align}
where $N$ is the e-folding number. 

The time evolution of the energy densities obtained by the numerical calculation is presented in Fig.~\ref{fig:sobolrhovst}. For most of the inflationary phase, the electric fields sustained by the quantum fluctuation are not amplified and the fermion production is negligible. 
A few e-folds before the inflation end, however, 
the electric fields rapidly grow and enter the attractor (see also Fig.~\ref{fig:EMvst}), and the Schwinger effect becomes active. After the inflation end, $\rho_{\psi}$ overwhelms $\rho_\phi$ and $\rho_E$.
Therefore, in this model, one observes the Schwinger preheating in which the fermions produced by the Schwinger effect dominate the universe after inflation. The oscillation of the electric fields is also seen as reported in Ref.~\cite{Gorbar:2019fpj}.

The Schwinger preheating is very fascinating, since it is a novel mechanism of preheating,
it may provide some inflation models with a new channel to reheat the universe, and it might change the thermal history.
Nevertheless, we do not know the generality of this phenomenon.
We investigate if the Schwinger preheating occurs in another model in the next section.


\subsubsection{The anisotropic inflation model}

		\begin{figure}
		\centering
		\includegraphics[width=.8\textwidth]{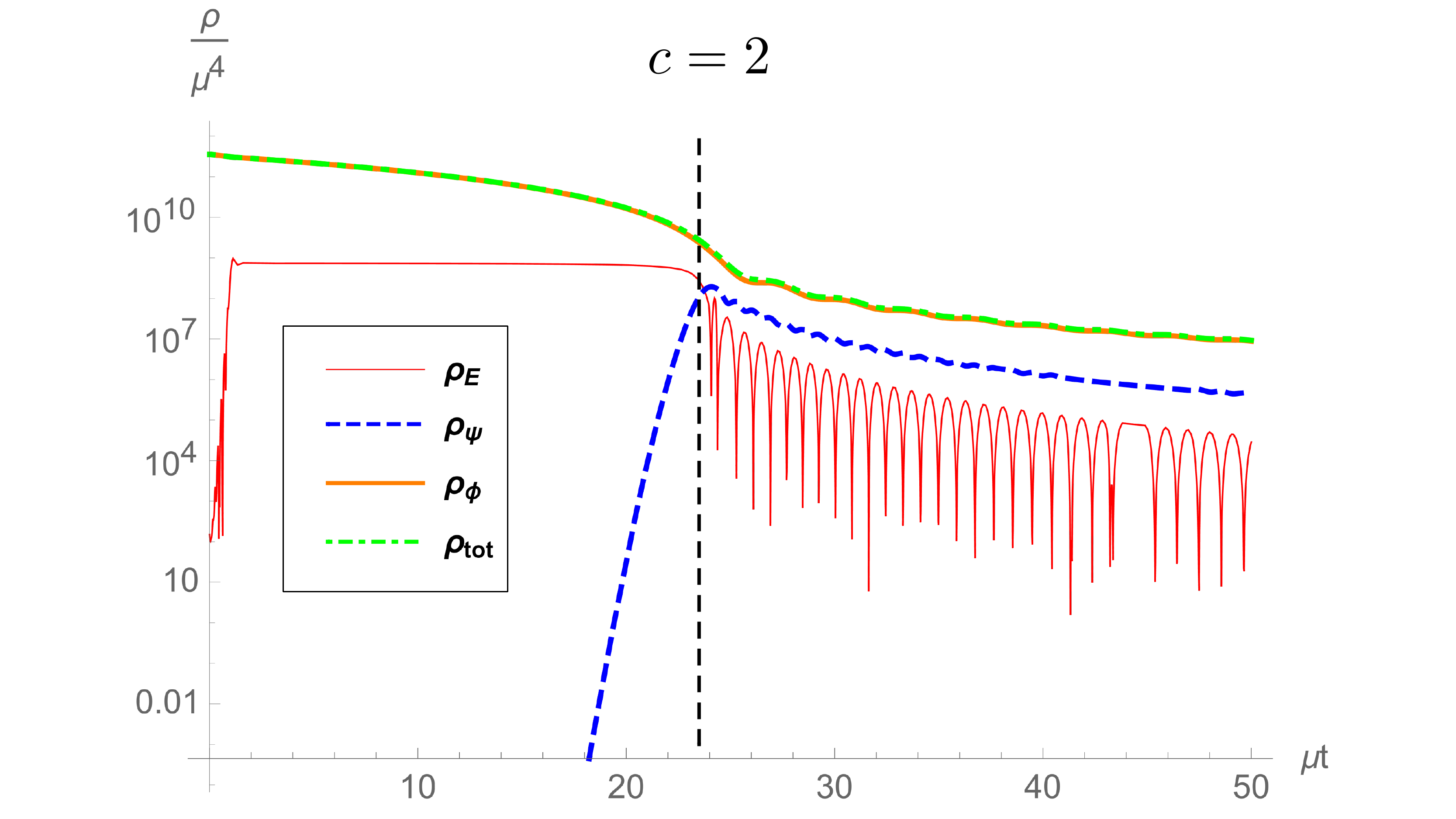}
		\caption{The numerical result in the case of the anisotropic inflation model $V_W$ and $f_W$ with the coupling constant $c=2$. The plot scheme is the same as Fig.~\ref{fig:sobolrhovst}. $\mu t=0$ corresponds to $N\simeq 56$ and inflation ends at $\mu t=23.5$. The Schwinger preheating does not occur in this case.
		}
		\label{fig:watanaberhovst}
	\end{figure}
As the second model, we investigate the anisotropic inflation model which was originally proposed in Ref.~\cite{Watanabe:2009ct}. Its potential and kinetic coupling have been given in Eq.~\eqref{eq:model of anisotropic inflation}.
The advantage of this model is that its dynamics has been well studied and some analytic solutions are available (e.g. Eq.~\eqref{eq:rhoE att Watanabe}).
Setting the initial condition as
\begin{align}
\phi(0)=\Mpl\sqrt{2N},\qquad
\phi'(0)=0,
\end{align}
we run the numerical calculation in the same way as the previous model.
The result is shown in Fig.~\ref{fig:watanaberhovst}. 
Although the electric fields enter the attractor much earlier than the previous model and the Schwinger effect lets $\rho_\psi$ exceed $\rho_E$ at around the inflation end, the produced fermions never dominate the universe. Thus, we find a counter-example of the Schwinger preheating. At least, the anistropic inflation model does not exhibit it for the coupling constant $c=2$.


	\begin{figure}
		\includegraphics[width=.45\textwidth]{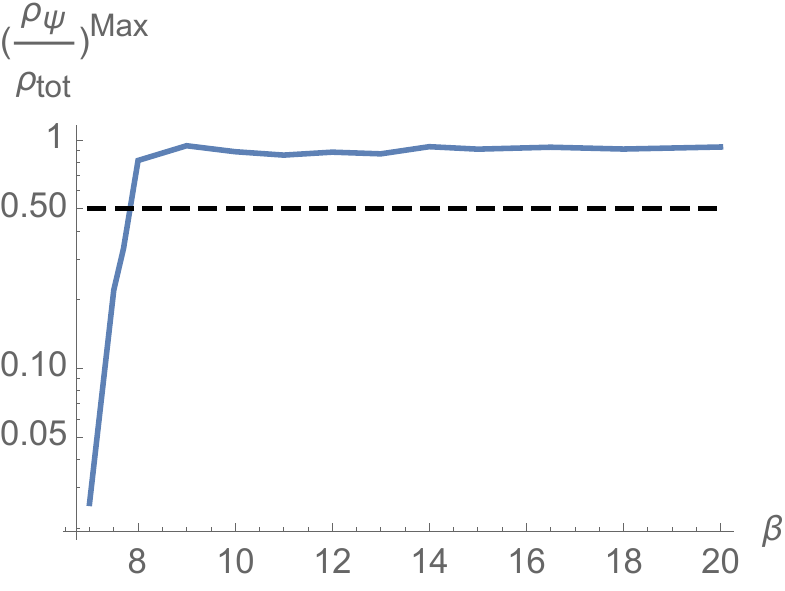}
			\hspace{5mm}
			\includegraphics[width=.45\textwidth]{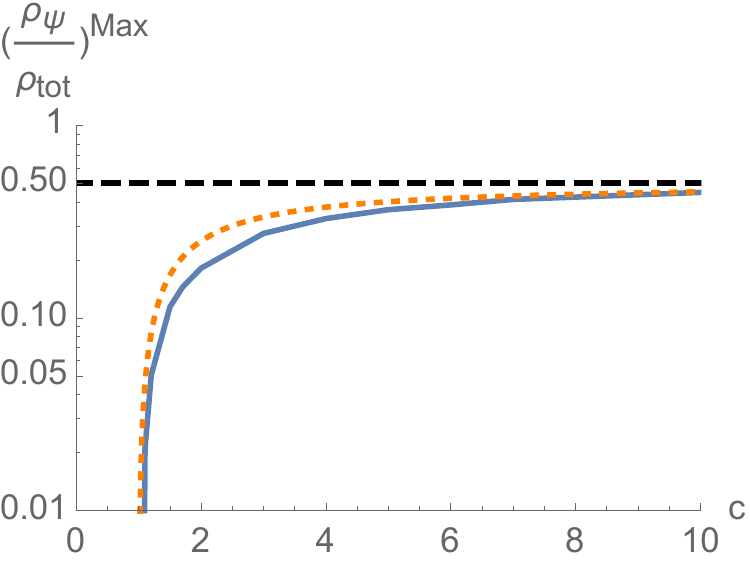}
			\caption{
			The maximum value of the energy fraction of the charged particles against the coupling constant. The left and right panels adopt the Starobinsky model with the Ratra coupling and the anisotropic inflation model, respectively. The black dashed line denotes the energy fraction equal to 1/2, and the Schwinger preheating is observed above this line. The orange dotted line in the right panel represents $(c-1)/2c$ which asymptotes to 1/2. The Schwinger preheating occurs only in the former model with the sufficiently strong coupling constant, $\beta\gtrsim 8$.}
			\label{fig:maximum energy fraction}
	\end{figure}
    It is naturally expected that the occurrence of the Schwinger preheating depends on the coupling constant.
    We show the dependence in the both models in Fig.~\ref{fig:maximum energy fraction}. When the maximum value of $\rho_\psi/\rho_{\rm tot}$ exceeds $1/2$, we observe the Schwinger preheating in the numerical calculations and vice versa. The anisotropic inflation model does not exhibit the Schwinger preheating, at least for $1<c\le 10$.
    Here, although one may think that even $\rho_\psi/\rho_{\rm tot}\approx 0.4$ is practically eligible for the Schwinger preheating, we adopt $1/2$ as the criteria.
    On the other hand, the Schwinger preheating occurs in the first model, if the coupling constant is large enough, $\beta \gtrsim 8$.
    Therefore, we find that the occurrence of the Schwinger preheating depends on models and coupling constants.
    We will further explore the dependence on the coupling constant in the next section.

		\section{Analytic Conditions for  the Schwinger Preheating}
			\label{sec:preheating}
		In this section, we analytically derive the conditions for the Schwinger preheating. 
		Since the Schwinger preheating occurs when the charged particles dominate the universe and $\rho_E$ supplies the energy to $\rho_{\psi}$ via the Schwinger effect, the necessary condition for the Schwinger preheating is $\rho_E \sim \rho_{\tot}$. Then we can obtain a model independent requirement for the Schwinger preheating by using the analytic expression for $\rho_E$, which is a function of $V(\phi)$ and $f(\phi)$. 
		
		Fortunately, this system is expected to have an attractor solution of the electric field for the following reason. The electric fields are amplified by the inflaton, because the inflaton's kinetic energy is transfer to the electric fields through the kinetic function. The amplification might be weaker than the dilution effect of the spacetime expansion in the early stage of the inflationary phase. Towards the end of inflation, however, the inflaton velocity $\dot{\phi}$ is accelerated by the potential force, the amplification effect is enhanced and eventually overcomes the dilution effect, and the electric fields begin to grow.  When the electric fields are sufficiently amplified and its backreaction on the inflaton becomes significant, the growth stops and $\rho_E$ is stabilized. This is because, if the electric fields are further amplified, the inflaton is slowed down by the backreaction, its kinetic energy pumped to the electric fields is reduced, and they decay. If the electric fields are smaller than the stabilized value, the inflaton is accelerated and gives more energy to the electric fields. Therefore, the inflaton and the electric fields constitute a negative feedback system and they should have an attractor solution.
		
		In what follows, we first derive the analytic expression for $\rho_E$ in the attractor phase as a function of $V$ and $f$. Second, we show that in the Starobinsky model with the Ratra coupling, once $\rho_E$ reaches the attractor, $\rho_E$ always dominates $\rho_{\tot}$ at the inflation end, regardless of the value of $\beta$.
		Finally we derive the condition for $\rho_E\sim \rho_{\rm tot}$.

		\subsection{The attractor solution for $\rho_E$}
		Here we find the general attractor solution for the energy density of the electric fields, $\rho_E^{\att}$. 
		As we discussed above, when the electric field reaches the attractor, its energy density is stabilized.
		Solving Eq.~\eqref{eq:EoM for rhoe an} without the fermions and the perturbations, we obtain
		\begin{equation}
		    e^{4\alpha}f^2 \rho_E = {\rm const}.
		 \label{eq:const rho}
		\end{equation}
		Thus, in the attractor phase, the kinetic function is expected to satisfy $f\propto e^{-2\alpha}$. 
		We rewrite this relation as
		\begin{align}
		\frac{df}{dt} = -2\dot{\alpha}f
		\quad \Longrightarrow \quad 
		\dot{\phi} = -2\frac{\dot{\alpha}}{(\ln f)'}.
		\label{eq: att phi}
		\end{align}
		Substituting it into Eq.~\eqref{eq:EoM for istoropic phi} and ignoring $\ddot{\phi}$ as a slow-roll correction term, we obtain
		\begin{align}
		\label{eq:attractor rhoe}
		\rho_E^{\att} = -\frac{V}{\Mpl^2(\ln f)'^2}+\frac{V'}{2(\ln f)'},
		\end{align}		
		where we used the slow-roll equaiton, $3\Mpl^2 \dot{\alpha}^2 \simeq V$.
		
		Now we confirm that the above solution, Eq.~\eqref{eq:attractor rhoe}, is the attractor solution by 
		studying the behavior of $\rho_E$ around $\rho_E^{\att}$.
		Considering a small deviation from the attractor solution, $\rho_E=\rho_E^{\att}+\delta\rho_E$,
		we rewrite the EoM for the inflaton, Eq.~\eqref{eq:EoM for istoropic phi}, as
		\begin{align}
		\delta\rho_E=\frac{3\dot{\alpha}}{2(\ln f)'^2}\left[2\dot{\alpha}+(\ln f)' \dot{\phi}\right].
		\end{align} 
		The EoM for $\rho_E$, Eq.~\eqref{eq:EoM for rhoe an}, reads $\dot{\rho}_E+2[2\dot{\alpha}+(\ln f)' \dot{\phi}]\rho_E=0$ in the isotropic background without the fermions.
		Substituting $\rho_E=\rho_E^{\att}+\delta\rho_E$ and ignoring $\dot{\rho}_E^{\att}$ as a slow-roll correction as well as the second order perturbation terms with $\delta\rho_E^2$, we obtain the evolution equation for $\delta\rho_E$ as
		\begin{align}
		\delta\dot{\rho}_E&=-\frac{4}{3}\frac{(\ln f)'^{2}\rho_E^{\att}}{\dot{\alpha}}\delta\rho_E.
		\end{align}
		This equation implies that $\delta\rho_E$ exponentially decays with a time scale, $\tau^{\att}=\frac{3\dot{\alpha}}{4(\ln f)'^{2}\rho_E^{\att}}$.
		Therefore, $\rho_E^{\att}$ is an attractor solution, when $\tau^{\att}$ is sufficiently short.  The comparison between $\tau^{\att}$ and the Hubble time $\dot{\alpha}^{-1}$ is given by
		\begin{align}
		\dot{\alpha}\tau^{\att}=\frac{\rho_{\tot}/\rho_E^{\att}}{4\Mpl^2(\ln f)'^{2}}= 2.5\times 10^{-2}\times
		\begin{cases}
        \left(\frac{\beta}{10}\right)^{-2}\left(\frac{\rho_E^{\att}/\rho_{\tot}}{10^{-1}}\right)^{-1} \, & \big(f = e^{\beta\frac{\phi}{\Mpl}}\big)\\
        c^{-2}\left(\frac{\phi}{10\Mpl}\right)^{-2}\left(\frac{\rho_E^{\att}/\rho_{\tot}}{10^{-1}}\right)^{-1} \, & \big(f = e^{\frac{c}{2}\left(\frac{\phi}{\Mpl}\right)^2}\big).
        \end{cases}
        \label{eq: Time scale}
		\end{align}
        Since $\dot{\alpha}\tau^{\att}$ is proportional to $\rho_{\rm tot}/\rho_E^{\att}$, this attractor is very weak for negligibly small electric fields, $\rho_E \ll \rho_{\rm tot}$. However, the attractor becomes stronger, as the electric fields are amplified and significantly backreact on the inflaton, as expected.

		In the case of the anistropic inflation model (i.e. $V=\mu^2\phi^2/2$ in the second line of Eq.~\eqref{eq: Time scale}),
		we can further reduce the time scale as
		$\dot{\alpha}\tau^{\att}=\frac{1}{4(c-1)}$, which is consistent with the known solution of $\rho_E$. 
		Thus, in the anisotropic inflation model, $\rho_E^{\att}$ is always the attractor solution irrespective of $\phi$, unless $c-1$ is negative or tiny.

		\begin{figure}
			\centering	
			\includegraphics[width=.45\textwidth]{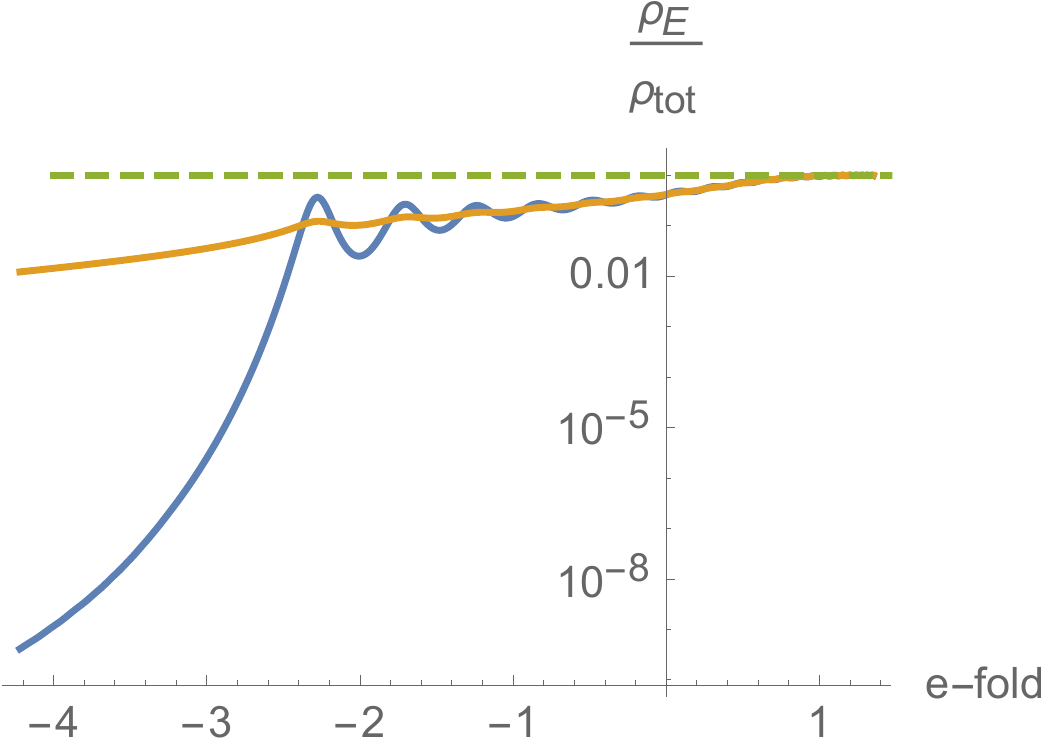}
			\hspace{5mm}
		    \includegraphics[width=.45\textwidth]{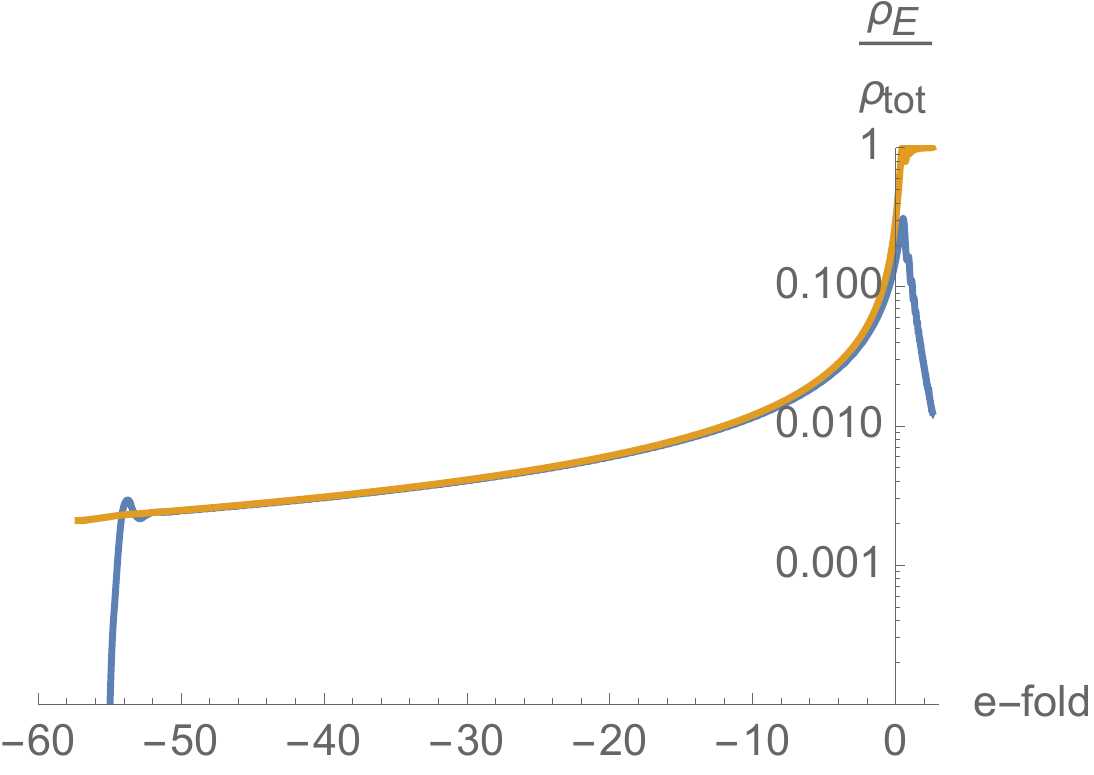}
			\caption{The energy density of the electric fields $\rho_E$ (blue) and that in the attractor phase $\rho_E^{\att}$ given in Eq.~\eqref{eq:attractor rhoe} (orange), which are normalized by the total energy density $\rho_{\rm tot}$. These numerical calculations do not include the fermions. The left and right panels adopt the Starobinsky model with the Ratra coupling for $\beta=10$ and the anisotropic inflation model for $c=2$, respectively. The horizontal axis denotes the e-folding number which is zero at the end of inflation ($\epsilon_H=1$) and takes negative values during inflation. In the left panel, $\rho_E$ does not reach the attractor until the last few e-folds of inflation, while it is well approximated by $\rho_E^{\att}$ even after inflation ends.
			In the right panel, on the other hand, $\rho_E$ stays at the attractor solution during inflation and deviates from it at around the end of inflation. When we include the fermions in the numerical calculation, the right panel does not change much, but in the left panel $\rho_E$ abruptly drops after the inflation end.
			}
			\label{fig:EMvst}
		\end{figure}
		We find that our attractor solution, Eq.~\eqref{eq:attractor rhoe}, matches the numerical results very well.  Fig.~\ref{fig:EMvst} presents the comparisons between the numerically computed $\rho_E$ and our analytic expression $\rho_E^{\att}$. Note that these numerical calculations do not include the fermion for illustrative purposes. The right panel shows that the electric fields stay at the attractor for most of the inflationary phase in the second model. However, the left panel shows that the electric fields enter the attractor only a few e-folds before the inflation end in the first model for $\beta=10$.
		Since $\dot{\alpha}\tau^{\att}\propto \beta^{-2}$ in the Ratra coupling case, for a smaller coupling constant, the attractor becomes weaker and the electric fields may not enter the attractor phase during inflation. 

\begin{figure}
			\centering	\includegraphics[width=.8\textwidth]{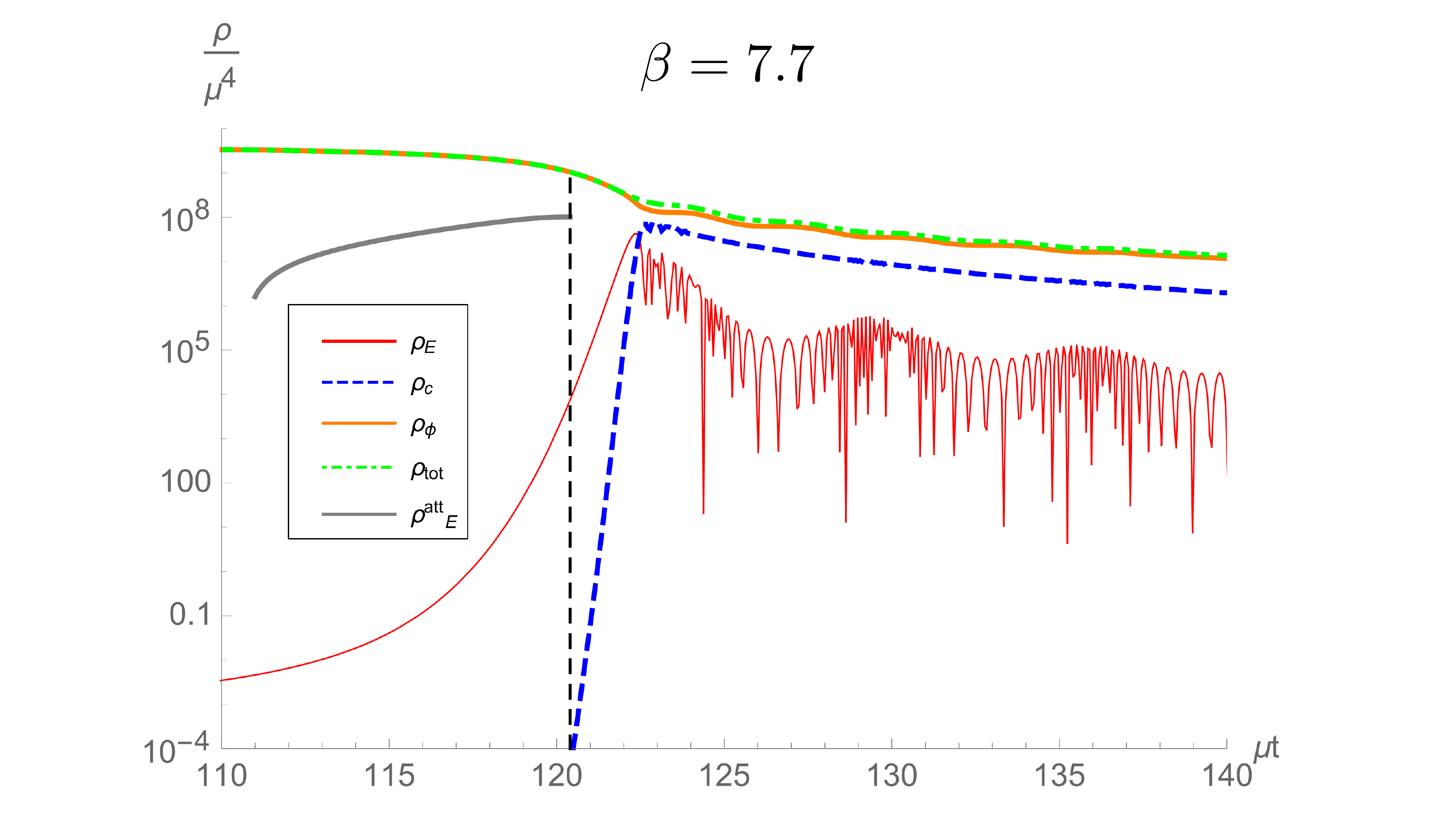}	
			\caption{ 
			The same plot as Fig.~\ref{fig:sobolrhovst} for a smaller coupling constant, $\beta=7.7$.
			The calculation starts at $\mu t=0$ corresponding to $N\simeq 60$ and inflation ends at $\mu t=120
			$ (vertical black dashed line). 
			The attractor solution $\rho_E^{\att}$ given in Eq.\eqref{eq:attractor rhoe} is shown as a black solid line until inflation ends. The electric fields do not enter the attractor phase during inflation. The maximum value of $(\rho_{E}+\rho_{\psi})/\rho_{\phi}$ is roughly equal to 1, and the Schwinger preheating is marginally absent.}
			\label{fig:sobolrhovst77}
		\end{figure}

        In Fig.~\ref{fig:sobolrhovst77}, we show the evolution of the energy densities in the first model
        for the coupling constant $\beta=7.7$, which is slightly smaller than the threshold value for the Schwinger preheating $\beta\approx 8$. In this case, the electric fields never enter the attractor
        and the Schwinger preheating does not occur. It infers that whether the electric fields reach the attractor solution is crucial to the Schwinger preheating.
		We will study this issue in more detail in Sec.~\ref{Condition to enter the attractor}.
		Before that, let us first investigate the cases where the electric fields enter the attractor.

	\subsection{The inflation end with the attractor}
	
	We have found that the Schwinger preheating occurs in the Starobinsky model with the Ratra coupling,
	if the coupling constant is large enough, $\beta \gtrsim 8$. In the same model, the electric fields
	enter the attractor during inflation for the same range of the coupling constant, $\beta \gtrsim 8$.
    Now we investigate the inflation end in the case that the electric fields are in the attractor phase. One finds that the slow-roll parameter is rewritten as
    \begin{align}
        \epsilon_H
        \equiv -\frac{\ddot{\alpha}}{\dot{\alpha}^2} = \frac{\dot{\phi}^2/2+2\rho_E/3}{\dot{\alpha}^2\Mpl^2}
        =\frac{2}{(\ln f)'^2\Mpl^2}+2\Omega_E,
        \qquad ({\rm attractor\ phase})
    \end{align}
    where $\dot{\phi}$ was evaluated with Eq.~\eqref{eq: att phi} which holds in the attractor phase, and $\Omega_E\equiv \rho_E/\rho_{\rm tot}$. The first and the second term in the R.H.S. represent
    the contributions from the inflaton kinetic energy and the electric energy density, respectively.
    As they grow, the slow-roll parameter $\epsilon_H$ increases and inflation eventually ends at $\epsilon_H=1$.
    
    
    In the case of the Ratra coupling $f_R$, we have $(\ln f)'=\beta/\Mpl$ and obtain
    \begin{align}
        \epsilon_H
        =\frac{2}{\beta^2}+2\Omega_E.
        \qquad ({\rm attractor\ phase\ \&\ Ratra\ coupling})
    \end{align}
    Since the first term is constant, $\Omega_E$ must be larger than one-half to end inflation.
    $\Omega_E$ does not have to stop increasing at $1/2$, and may become even larger after inflation,
    until the electric fields are significantly consumed to produce and accelerate the fermions.
    This is a general property of inflationary models with the Ratra coupling in the attractor phase of the electric fields.
    
    The above argument apparently suggests that the models with Ratra coupling always have a large $\Omega_E$ at the inflation end and lead to the Schwinger preheating. However, as shown in Fig.~\ref{fig:maximum energy fraction}, the Starobinsky model with the Ratra coupling does not cause the Schwinger preheating for $\beta \lesssim 8$. This is because, for a smaller coupling constant, the electric fields do not reach the attractor solution, as seen in Fig.~\ref{fig:sobolrhovst77}.

	\subsection{Condition to enter the attractor}
	\label{Condition to enter the attractor}
	
		Entering the attractor phase of the electric fields is a requirement to cause the Schwinger preheating in the Starobinsky model with the Ratra coupling. In this section, we will derive the condition to reach the attractor solution.

		Let us introduce the inflaton value $\phi^\ast$ at which the electric fields begin to grow by overcoming the cosmic dilution.
		At this moment, the kinetic function transiently satisfies $\dot{\phi}^\ast=-2\dot{\alpha}/(\ln f)'=-2\dot{\alpha}\Mpl/\beta$ for the same reason as Eq.~\eqref{eq: att phi}. 
		Using the slow-roll approximation, we obtain $\phi^\ast$ as
		\begin{align}
		\phi^{\ast}&=\sqrt{\frac{3}{2}} \log\left(1+\frac{\sqrt{6}}{3}\beta\right)\Mpl.
		\end{align}
		At $\phi=\phi^\ast$, the energy density of the electric field sourced by the perturbation is also obtained as 
		\begin{equation}
		    \dot{\rho}_E+8\dot{\alpha}\rho_E =\frac{5\dot{\alpha}^5}{4\pi^2},
		    \quad\Longrightarrow\quad
		    \rho_E(\phi^{\ast})=\frac{5\dot{\alpha}^{4}_\ast}{32\pi^2}=\frac{5V^2(\phi^{\ast})}{288\pi^2\Mpl^4}.
		\end{equation}
		Here we ignored the anisotropy and the fermions, and used $\dot{f}/f=2\dot{\alpha}$ in Eq.~\eqref{eq:EoM for rhoe an}.
		%
		After $\phi=\phi^\ast$, from Eq.~\eqref{eq:const rho}, the electric energy density is given by
		\begin{align}
		\label{eq:exact rhoE}
		\rho_E(\phi)=\frac{5V^2(\phi^{\ast})}{288\pi^2\Mpl^4}
		\exp\left[2\beta(\phi^{\ast}-\phi)+4N\right],
		\end{align}
		where we introduced the e-folding number which is computed as,
		\begin{align}
		N=\int_{t^{\ast}}^tH dt= \int^{\phi^{\ast}}_{\phi}\frac{V}{V'}d\phi=\left[\frac{3}{4}e^{\sqrt{\frac{2}{3}}\phi}-\sqrt{6}\phi\right]^{\phi^{\ast}}_{\phi}.
		\end{align}
		Now we can calculate when $\rho_E(\phi)$ reaches $\rho_E^{\att}$.
		Numerically solving
		\begin{align}
		\rho_E(\phi^{\att})=\rho^{\att}_E(\phi^{\att}),
		\label{eq:phi att euqation}
		\end{align}
		we obtain the inflaton value $\phi^{\att}(\beta)$, at which the electric fields enter the attractor phase, as a function of $\beta$.
		Finally, we impose a condition that the electric fields enter the attractor before the end of inflation at $\phi=\phi_{\rm end}$,
		\begin{align}
		\phi^{\att}(\beta)>\phi_{\rm end}.
		\end{align}
		
		\begin{figure}
			\centering
			\includegraphics[width=.7\textwidth]{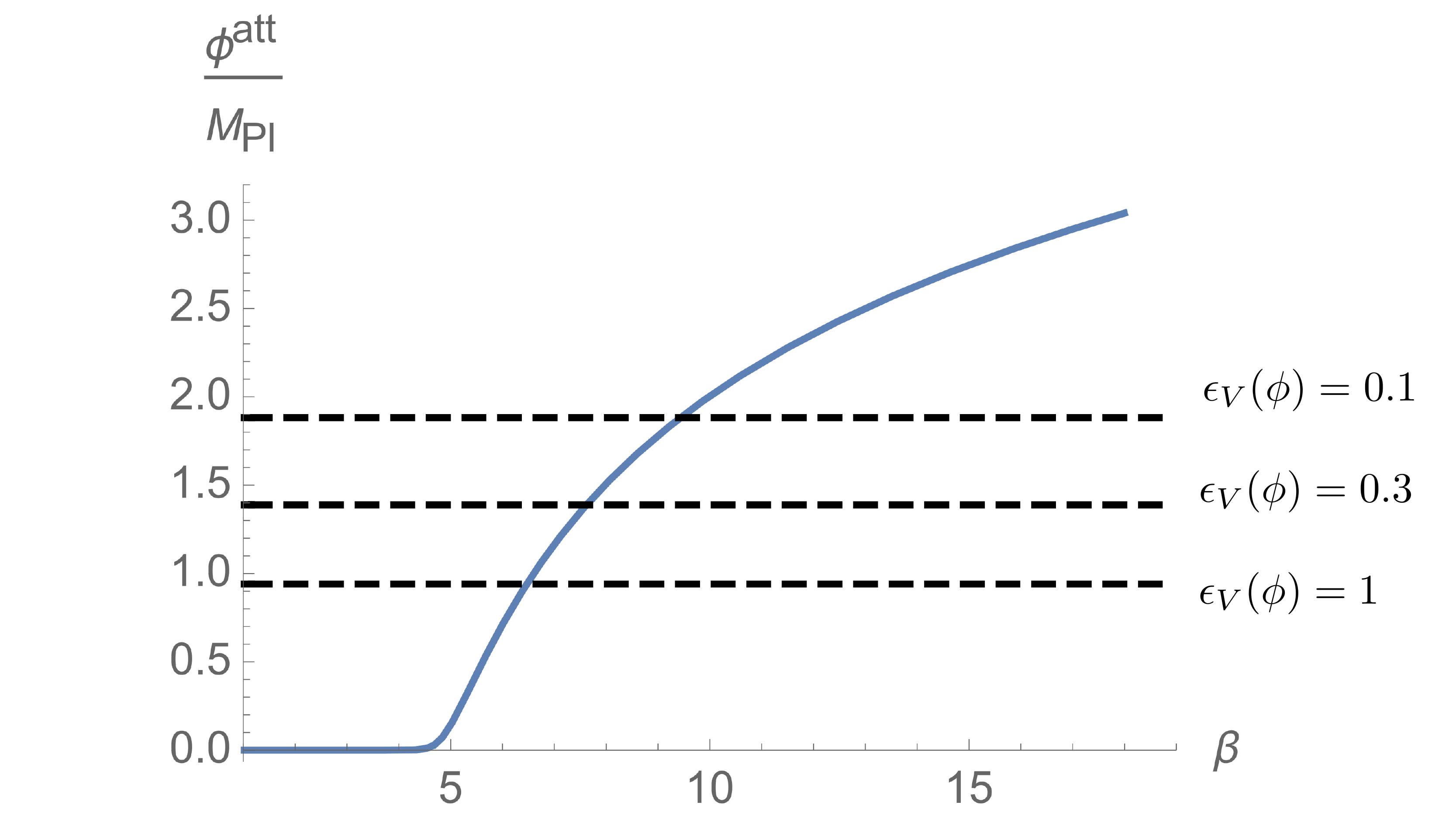}
			\caption{ The inflaton value $\phi^{\att}(\beta)$ at which the electric fields enter the attractor phase and which solves Eq.~\eqref{eq:phi att euqation}. 
			For the electric fields to enter the attractor during inflation, $\phi^{\att}(\beta)$ must be larger than $\phi_{\rm end}$, which sets the lower bound on $\beta$. The dashed lines represent the values of $\phi$ satisfying $\epsilon_V(\phi)=1, 0.3$ and $0.1$ from bottom to top, and they cross $\phi^{\att}(\beta)$ at $\beta= 6.6,  7.6$ and $9.4$}
			\label{fig:attcond}
		\end{figure}
		In Fig.~\ref{fig:attcond}, we plot $\phi^{\att}(\beta)$ and compare it with the inflaton values satisfying $\epsilon_V(\phi)=0.1, 0.3$ and 1.
		If one adopt $\epsilon_V(\phi_{\rm end})=1$ to characterize the inflation end, one obtains $\beta=6.5$ as the threshold value, which is a slightly milder condition than the numerical result shown in Fig.~\ref{fig:maximum energy fraction}.
		Note that since we extrapolated the dynamics around the end of inflation by using the expressions derived under the slow-roll approximation, we may well have a residual error. In fact, when one uses $\epsilon_V(\phi_{\rm end})=0.3$ instead, the threasold would be $\beta=7.6$ which is closer to the numerical result.

        Consequently, we found that the occurrence of the Schwinger preheating in the Starobinsky model with the Ratra coupling is determined by whether the electric fields enter the attractor solution during inflation or not.
        For a weaker coupling than $\beta=8$, the amplification of the electric fields is too slow to reach the attractor by the end of inflation, and the charged fermions cannot receive a sufficiently large energy to dominate the universe.
		

	   \section{Influence of the Anisotropy}
	   \label{sec:anisotropy}
	
	We have neglected the spacetime anisotropy since Eq.~\eqref{eq:eom for jcond}. However, the isotropic spacetime is inconsistent with the background electric field, when it is strong enough to significantly contribute to the spacetime evolution. In this case, 
	we should work in the anisotropic background, Eq.~\eqref{eq:Bianchi-I}. In this section, we restore the anisotropy and investigate its influence on the evolution of the matter fields.
	
	    
	    In Eqs.~\eqref{eq:eom for n}-\eqref{eq:eom for jcond}, we wrote the hydrodynamic equations in the anisotropic background, while we left $\mathcal{J}_{xx}^{(2, -2)}$ and $\mathcal{J}_{xxx}^{(3, -2)}$ unspecified, which became irrelevant in the isotropic limit $\dot{\sigma}\to 0$.
	    Here let us determine their expressions through phenomenological arguments in similar ways to $\mathcal{J}_{xx}^{(2,0)}$. Assuming 
	    $j_{\rm cond}=n v_x$ and combining Eq.~\eqref{eq:eom for n} and \eqref{eq:eom for jcond}, we obtain the EoM for the charged particles as
		\begin{align}
		\label{eq: derude1}
		\frac{d v_x}{dt}=-\frac{v_x}{n}\left(2g\Gamma+\frac{n_{\rm eq}}{\tau}\right)+\frac{\mathcal{E}}{n}\mathcal{J}_{\perp\perp}^{(2,0)}+3\dot{\sigma}\left(v_x-\frac{\mathcal{J}_{xxx}^{(3,-2)}}{n}\right).
		\end{align}
		Compared to the istropic case, Eq.~\eqref{eq: derudes1}, the anistropy induces the last term in the R.H.S.\footnote{Since we took the massless limit, the contribution from the isotropic expansion vanishes.}
		The EoM for the relativistic charged particles in the anisotropic background \eqref{eq:Bianchi-I} under the electric fields in the $x$-direction is written as
		\begin{align}
		\label{eq: derude2}
		\frac{dv_x}{dt}=-Cv_x+\mathcal{E}\frac{n}{\rho_{\psi}}(1-v_x^2)+3\dot{\sigma}(v_x-v_x^3).
		\end{align}
		Comparing Eq.~\eqref{eq: derude1} with Eq.~\eqref{eq: derude2}, we obtain an additional closure condition as
		\begin{align}
		\mathcal{J}_{xxx}^{(3,-2)}&=j_{\rm cond}\left(\frac{j_{\rm cond}}{n}\right)^2.
		\end{align}
		In the meanwhile, the same expressions for $\mathcal{J}_{xxx}^{(3,-2)}$ as well as $\mathcal{J}_{\perp\perp}^{(2,0)}$ can be obtained by approximately manipulating their definitions, 
		\begin{align}
		\mathcal{J}_{\perp\perp}^{(2,0)}&\equiv n\left<\frac{p_{\perp}^2}{\epsilon_{\bm{p}}^3}\right>\simeq n\frac{1-\braket{\frac{p_{\parallel}}{\epsilon_{\bm{p}}}}^2}{\left<\epsilon_{\bm{p}}\right>}= \frac{n^2-j_{\rm cond}^2}{\rho_{\psi}},
		\notag\\
		\mathcal{J}_{xxx}^{(3,-2)}&\equiv n\left<\left(\frac{p_{\parallel}}{\epsilon(\bm{p})}\right)^3\right>
		\simeq n\left<\frac{p_{\parallel}}{\epsilon(\bm{p})}\right>^3
		=j_{\rm cond}\left(\frac{j_{\rm cond}}{n}\right)^2,
		\end{align}
		where we used $v_x=\braket{\frac{p_{\parallel}}{\epsilon_{\bm{p}}}}$, $\rho_c=n\braket{\epsilon_{\bm{p}}}$, and the bracket $\braket{\cdots}$ denotes the average over the phase space weighted by $\mathcal{F}(t,\bm{p})$. 
		Although the phenomenological argument to identify $\mathcal{J}_{xx}^{(2,-2)}$ is unknown,
		we can apply a similar approximation to find its expression as
		\begin{align}
		\mathcal{J}_{xx}^{(2,-2)}\equiv n\left<\frac{p_{\parallel}^2}{\epsilon}\right>\simeq n\left<\epsilon\right>\left<\frac{p_{\parallel}}{\epsilon}\right>^2=\rho_{\psi}\left(\frac{j_{\rm cond}}{n}\right)^2.
		\end{align}
		Without the verification from the microscopic EoM, one may be concerned with 
		the accuracy of the closure condition for $\mathcal{J}_{xx}^{(2,-2)}$.
		Nevertheless, since the term with $\mathcal{J}_{xx}^{(2,-2)}$ evaluated by the above expression is always subdominant in Eq.~\eqref{eq:eom for rhoc} in our numerical calculations, a small error in
		the expression, if any, would not affect our final results.
		
		
		
			\begin{figure}
				\centering	\includegraphics[width=.45\textwidth]{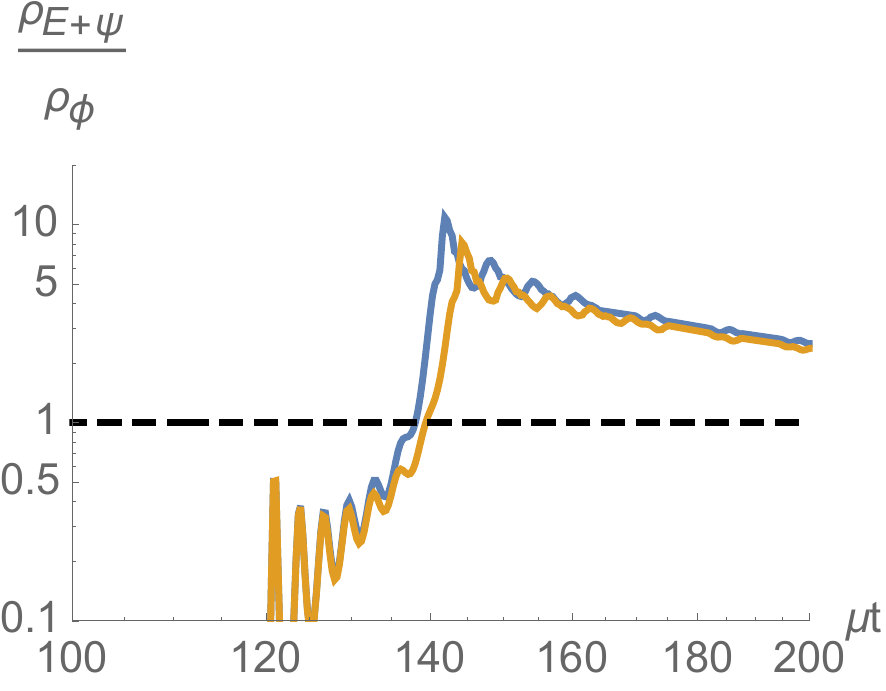}
				\hspace{5mm}	\includegraphics[width=.45\textwidth]{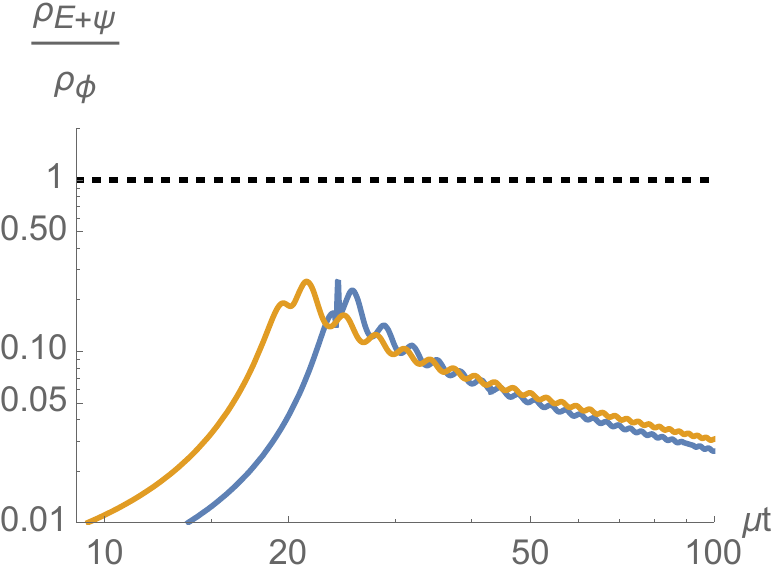}
				\caption{The time evolution of the energy fraction $(\rho_E+\rho_\psi)/\rho_\phi$ computed in the anisotropic (blue) and the isotropic background (orange).
				The left and right panels adopt the Starobinsky model with the Ratra coupling for $\beta=10$ and the anisotropic inflation model for $c=2$, respectively.
				The background anisotropy makes the peak height and its timing of the energy fraction slightly higher and earlier in the left panel, respectively, while it is opposite (i.e. slightly lower and later) in the right panel.
				}
				\label{fig:anivsi}
			\end{figure}
		We obtain the full hydrodynamic equations for this system, and we can perform numerical calculations in the anisotropic background. 
		We consider the same models as Sec.~\ref{sec:preheating}, which are the Starobinsky model with the Ratra coupling, Eq,~\eqref{eq:starobinskyratra}, and the anisotropic inflation model, Eq~\eqref{eq:model of anisotropic inflation}. The numerical results in the anisotropic and isotropic background are compared in Fig.~\ref{fig:anivsi}. The anisotropy does not have a dramatic impact on the evolution of the energy densities.
	    One observes that $(\rho_E+\rho_\psi)/\rho_\phi$ is slightly shifted by the anisotropy, while the overall behavior stays almost the same in the both models.

		\section{Summary and Discussion}
		\label{sec:Summary and Discussion}
	    
	    In this paper, we applied the kinetic hydrodynamic approach to study the Schwinger effect 
	    in the Starobinsky inflation model with the Ratra coupling and the anisotropic inflation model. The former model was analyzed in the previous work~\cite{Gorbar:2019fpj}, and they concluded that the Schwinger preheating occurred when the coefficient of the Ratra coupling satisfies $\beta\geq 8$ without any analytical discussion. 
	    While we confirmed the results of the previous work, we have shown that the Schwinger preheating does not occur in the latter model. 
	    
	    We also analytically discussed the condition for the Schwinger preheating.
	    The Schwinger preheating occurs when the energy density of the charged particles dominates the universe and the energy density of the charged particle is supplied from electric fields. Thus the necessary conditions for the Schwinger preheating is $\rho_E/\rho_{\tot}\sim 1/2$, which leads to the significant backreaction from the electric fields on the inflaton. We found the general attractor solution of $\rho_E$ by using the slow-roll approximation, and investigated the stability of the attractor solutions. 
	    Our result is consistent with the conventional analysis of the anisotropic inflation model. Using the attractor solution, we evaluated the electric field energy density at the inflaton end. 
	    In the Starobinsky model with the Ratra coupling, the analytic evaluation apparently indicates that $\rho_E$ always dominates the universe at the end of inflation, independent of $\beta$.
	    However, 
	    we found that $\rho_E$ does not reach the attractor, and hence it remains much smaller than $\rho_{\phi}$ in the case of a small $\beta$. The threshold to enter the attractor is estimated as $\beta=7.5$ when the inflation end is evaluated as $\epsilon_H=0.3$, while the numerical results indicate the threshold is $\beta\approx 8$. 
	    
	    We have also addressed the impact of the spacetime anisotropy. Two more closure conditions to cut the chain equation are needed, and we specified them. However we still do not know the physical meaning of one of the two closure conditions. Although the energy fraction of the electric fields approaches 1/2 and a significant effect of the anisotropy on the background evolution could be expected, the numerical results showed that the dynamics is almost same as the isotropic case except for the slight shifts of the timing of the Schwinger preheating and the maximum value of $\left(\rho_E+\rho_{\psi}\right)/\rho_{\phi}$. We found that the anisotropy hastens the preheating in Starobinsly and Ratra model but delays it in anisotropic inflation. The origin of these difference is beyond the scope of this paper. 
	    
	    The Schwinger preheating may give a new reheating channel to inflation models and change the thermal history of the universe. Hence it is meaningful to investigate its dynamics and find its condition. 
	    In our analysis, we do not consider a comprehensive model including inflaton decay processes. If the inflaton itself or its decay products are coupled to the charged fermions produced by the Schwinger effect, for instance, the Schwinger preheating might trigger the reheating. It would be interesting to introduce these couplings and study the dynamics after the Schwinger preheating. 
	    We hope to come back to these issues in the future work.
	    
	    The quantum kinetic approach is well studied approach to the Schwinger effect in the case of the spatially uniform dynamical electric fields. According to Ref.~\cite{Schmidt:1998zh}, the general form of the Schwinger source term includes the integration of the distribution function in time except for the case of low number density of fermions, which is the same situation as our case.
	    Since the deviation from the thermal equilibrium is not always small in the low number density, the low number density approximation is possibly in conflict with the ansatz where the collision term is described by the restoring force. 
	    In order to the more general analysis, we should relax either one of the two assumptions. 
	    We will investigate the other prescription to deal the kinetic approach in the future work.
	    
\acknowledgments
We would like to thank Keigo Shimada, Shin Kobayashi, Shintaro Sato and Teruaki Suyama for useful discussions and comments.
The work of T.F. was supported by JSPS KAKENHI No.~17J09103 and No.~18K13537.

	\appendix
		\section{Derivation of the hydrodynamic Boltzmann equation}
		\label{appendixA}

		Let us obtain the Eq.~\eqref{eq:hydro boltzmann an} from the Boltzmann equation in the Bianchi-I spacetime, Eq.~\eqref{eq:Boltzmann an}.
		\subsection*{The first term in Eq.~\eqref{eq:Boltzmann an}}
		Since the moment of the distribution function is independent from the momentum, the partial time derivative for $\mathcal{F}_{\bm{p}}$ becomes a total derivative as,
		\begin{align}
		2g\int \frac{d^3\bm{p}}{(2\pi)^3} \frac{p_{i_1}\cdots p_{i_k}}{\epsilon(\bm{p})^{r+2k-1}}\frac{\partial\mathcal{F}_{\bm{p}}}{\partial t}=\frac{d}{dt}\mathcal{J}^{(k,r)}_{i_1,\cdots,i_{k}}.
		\end{align}
		\subsection*{The second term in Eq.~\eqref{eq:Boltzmann an}}
		By using the integration by part, the term proportional to $\mathcal{E}(t)$ is written as
		\begin{align}
		2g\int \frac{d^3\bm{p}}{(2\pi)^3} \frac{p_{i_1}\cdots p_{i_k}}{\epsilon(\bm{p})^{r+2k-1}}\frac{\partial\mathcal{F}_{\bm{p}}}{\partial p_{\parallel}}=(r+2k-1)\mathcal{J}^{(k+1,r)}_{i_1,\cdots i_k,x}-\sum_{\ell=1}^{k}\delta_{i_{\ell},x}\mathcal{J}^{(k+1,r+2)}_{i_1,\cdots,\hat{i}_ {i_{\ell}},\cdots,x},
		\end{align}
		where we used the relation,
		\begin{align}
		\epsilon(\bm{p})&=\sqrt{\bm{p}^2+m^2},\\
		\frac{\partial}{\partial p_{\parallel}}\left(\epsilon(\bm{p})^{-r-2k+1}\right)&=-(r+2k-1)\epsilon(\bm{p})^{-r-2k-1}p_{\parallel}.
		\end{align}
		In the same way as $\mathcal{E}(t)$ term, the term proportional to $3\dot{\sigma}$ is written as,
		\begin{align}
		2g\int \frac{d^3\bm{p}}{(2\pi)^3} \frac{p_{i_1}\cdots p_{i_k}}{\epsilon(\bm{p})^{r+2k-1}}p_{\parallel}\frac{\partial\mathcal{F}_{\bm{p}}}{\partial p_{\parallel}}&=-2g\int \frac{d^3\bm{p}}{(2\pi)^3} \frac{\partial}{\partial p_{\parallel}}\left(\frac{p_{i_1}\cdots p_{i_k}p_{\parallel}}{\epsilon(\bm{p})^{r+2k-1}}\right)\mathcal{F}_p,
		\notag\\
		&=(r+2k-1)\mathcal{J}_{i_1\cdots i_kxx}^{(k+2,r-2)}-\mathcal{J}_{i_1\cdots i_k}^{(k,r)}-\sum_{\ell=1}^{k}\mathcal{J}_{i_1\cdots \hat{i_{\ell}}\cdots i_k x}^{(k,r)},
		\end{align}
		where we used the following relation,
		\begin{align}
		\frac{\partial}{\partial p_{\parallel}}\left(\frac{p_{i_1}\cdots p_{i_k}p_{\parallel}}{\epsilon(\bm{p})^{r+2k-1}}\right)=-(r+2k-1)\frac{p_{i_1}\cdots p_{i_k}p_{\parallel}^2}{\epsilon(\bm{p})^{r+2k+1}}+\sum_{\ell=1}^k\delta_{\ell,x}\frac{p_{i_1}\cdots \hat{p}_{i_{\ell}}\cdots p_{i_k}p_{\parallel}}{\epsilon(\bm{p})^{r+2k-1}}+\frac{p_{i_1}\cdots p_{i_k}}{\epsilon(\bm{p})^{r+2k-1}}.
		\end{align}
		\subsection*{The third term in Eq.~\eqref{eq:Boltzmann an}}
		The term proportional to the $(\dot{\alpha}+\dot{\sigma})$ is written as,
		\begin{align}
		2g(\dot{\alpha}+\dot{\sigma})\int \frac{d^3\bm{p}}{(2\pi)^3}\frac{p_{i_1},\cdots,p_{i_k}}{\epsilon(\bm{p})^{r+2k-1}}\bm{p}\cdot\frac{\partial \mathcal{F}_{p}}{\partial\bm{p}}=-(r+2k-1)(\dot{\alpha}+\dot{\sigma})m^2\mathcal{J}^{(k,r+2)}_{i_1,\cdots,i_k}-(4-k-r)(\dot{\alpha}+\dot{\sigma})\mathcal{J}^{(k,r)}_{i_1,\cdots,i_{k}},
		\end{align}
		where we used the relations,
		\begin{align}2g(\dot{\alpha}+\dot{\sigma})\int \frac{d^3\bm{p}}{(2\pi)^3}\frac{p_{i_1},\cdots,p_{i_k}}{\epsilon(\bm{p})^{r+2k-1}}\bm{p}\cdot\frac{\partial \mathcal{F}_{p}}{\partial\bm{p}}=-2g(\dot{\alpha}+\dot{\sigma})\int \frac{d^3\bm{p}}{(2\pi)^3}\frac{\partial}{\partial\bm{p}}\left(\frac{p_{i_1}\cdots p_{i_{k}}\bm{p}}{\epsilon(\bm{p})^{r+2k-1}}\right)\mathcal{F}_{p},
		\end{align}
		and
		\begin{align}
		\frac{\partial}{\partial\bm{p}}\left(\frac{p_{i_1}\cdots p_{i_{k}}\bm{p}}{\epsilon(\bm{p})^{r+2k-1}}\right)&=(k+3)\frac{p_{i_1}\cdots p_{i_k}}{\epsilon(\bm{p})^{r+2k-1}}-(r+2k-1)|\bm{p}|^2\frac{p_{i_1}\cdots p_{i_k}}{\epsilon(\bm{p})^{r+2k+1}},\notag\\
		&=(k+3)\frac{p_{i_1}\cdots p_{i_k}}{\epsilon(\bm{p})^{r+2k-1}}-(r+2k-1)(\epsilon^2-m^2)\frac{p_{i_1}\cdots p_{i_k}}{\epsilon(\bm{p})^{r+2k+1}},\notag\\
		&=-(r+k-4)\frac{p_{i_1}\cdots p_{i_k}}{\epsilon(\bm{p})^{r+2k-1}}+(r+2k-1)m^2\frac{p_{i_1}\cdots p_{i_k}}{\epsilon(\bm{p})^{r+2k+1}}.
		\end{align}
Thus we obtain Eq.~\eqref{eq:hydro boltzmann an} by using these discussions. 


\end{document}